\documentclass[pra,twocolumn,superscriptaddress,showpacs,floatfix]{revtex4}
\usepackage{hyperref}
\usepackage{lipsum}
\usepackage[british]{babel}
\usepackage[utf8]{inputenc}
\usepackage{amsmath}
\usepackage{amssymb}
\usepackage{mathtools}
\usepackage{braket}
\usepackage{graphicx}
\usepackage{xcolor}
\usepackage{braket}

\DeclareMathAlphabet{\mathpzc}{OT1}{pzc}{m}{it}

\begin{document}

\title{An intuitive picture of the Casimir effect}
\author{Daniel Hodgson}
\affiliation{The School of Physics and Astronomy, University of Leeds, Leeds LS2 9JT,
United Kingdom}
\author{Christopher Burgess}
\affiliation{The School of Physics and Astronomy, University of Leeds, Leeds LS2 9JT,
United Kingdom}
\author{M. Basil Altaie}
\affiliation{Department of Physics, Yarmouk University, 21163 Irbid, Jordan}
\author{Almut Beige}
\affiliation{The School of Physics and Astronomy, University of Leeds, Leeds LS2 9JT,
United Kingdom}
\author{Robert Purdy} 
\affiliation{The School of Physics and Astronomy, University of Leeds, Leeds LS2 9JT,
United Kingdom}
\date{\today}

\begin{abstract}
The Casimir effect, which predicts the emergence of an attractive force between two parallel, highly reflecting plates in vacuum, plays a vital role in various fields of physics, from quantum field theory and cosmology to nanophotonics and condensed matter physics. Nevertheless, Casimir forces still lack an intuitive explanation and current derivations rely on regularisation procedures to remove infinities. Starting from special relativity and treating space and time coordinates equivalently, this paper overcomes no-go theorems of quantum electrodynamics and obtains a local relativistic quantum description of the electromagnetic field in free space. When extended to cavities, our approach can be used to calculate Casimir forces directly in position space without the introduction of cut-off frequencies.
\end{abstract}

\maketitle

\section{Introduction}

Since its initial discovery, the Casimir effect \cite{Casimir}, which predicts an
attractive force between two highly reflecting plates in vacuum, has received a lot of attention in the
literature \cite{interest}. Despite not having a counterpart in classical
electrodynamics, recent experiments confirm its predictions \cite{exps}. Nevertheless, there is still some controversy surrounding the origin of this effect \cite{interest2}.  For example, the standard derivation requires regularisation procedures before identifying a finite contribution to the zero point energy of the electromagnetic (EM) field which depends on the mirror distance $D$ \cite{Milonni}. Moreover, the standard derivation simply assumes that the plates restrict the quantised EM field inside the cavity to standing waves with a discrete set of so-called resonant frequencies. However, standing wave mode models cannot take into account from which direction light enters an optical cavity and therefore cannot reproduce the typical behaviour of Fabry-Perot cavities which have maximum transmission rates for resonant light \cite{Barlow}. Discrete mode models also imply that no light is permitted inside a cavity with mirror distances well below typical optical wavelengths, which contradicts recent experiments with nanocavities \cite{Baumberg}. To fully capture all aspects of optical cavities, an alternative approach is needed.

The common view of the quantised EM field as a collection of energy quanta with well-defined wave vectors $k$, positive frequencies $\omega = c|k|$, energies $\hbar \omega$ and polarisations $\lambda = {\sf H},{\sf V}$ can be traced back to Planck's 1901 modelling of black body radiation \cite{Planck} and Einstein's 1917 analysis of the photoelectric effect \cite{Einstein}. Using a canonical quantization prescription, textbooks usually obtain expressions for the basic field observables by expanding the vector potential $\boldsymbol{A}$ of the classical EM field into a Fourier series. Its coefficients are then replaced by photon creation and annihilation operators with bosonic commutation relations \cite{Milonni,Sakurai,Loudon}. For light propagating along the $x$ axis, these excitations describe wave packets travelling with the speed of light $c$ in a well-defined direction \cite{Bennett}. Despite being relativistic, they evolve according to a Schr\"odinger equation with a Hamiltonian which coincides with their positive energy observable.

Within the above formalism, it has been challenging to establish a local theory of light without ambiguities. 
Several no-go theorems have been put forward regarding the localisability of its elementary particles \cite{Pryce}. The main contributor to these difficulties is the lack of a position operator for the photon, which leaves no clear candidate for a single photon wave function \cite{BB2,Sipe,Raymer,LandauandPeierls}. In spite of these complications, a local description of the EM field would lend itself naturally to the modelling of locally-interacting quantum systems and other quantum optics experiments \cite{Axel,linopt}. Consequently a lot of effort has been made to introduce more practical notions of localisability and, in some cases, make alterations to the current formalism such that localisation becomes possible (cf.~e.g.~\cite{Mandel,Cook1,Hawton1,Jake} and Refs.~therein).  

Recently, we introduced a description of the quantised EM field in terms of local annihilation and creation operators, $a_{s\lambda }(x,t)$ and $a^\dagger_{s\lambda }(x,t)$, in the Heisenberg picture with $x$ and $t$ denoting the spacetime coordinates of local field excitations and with $s=\pm 1$ and  $\lambda ={\sf H}, {\sf V}$ indicating the direction of propagation and the polarisation of the associated field vectors  \cite{Jake}. To overcome the above localisability issues, we had to double the Hilbert space of the quantised EM field and allow for positive- and negative-frequency solutions in its momentum-space representation (see also Refs.~\cite{Cook1,Hawton1,Rubino,Bianca}). There now appear to be two different types of Hamiltonians: the positive energy observable $H_{\rm energy}$ and the dynamical Hamiltonian $H_{\mathrm{dyn}}$ which has positive and negative eigenvalues and governs the dynamics of wave packets. As it has been previously noted by other authors, a dynamical Hamiltonian which is not bounded from below is needed to guarantee causality \cite{Hegerfeldt}. 

\begin{figure*}[t]
\centering
\includegraphics[width=\textwidth]{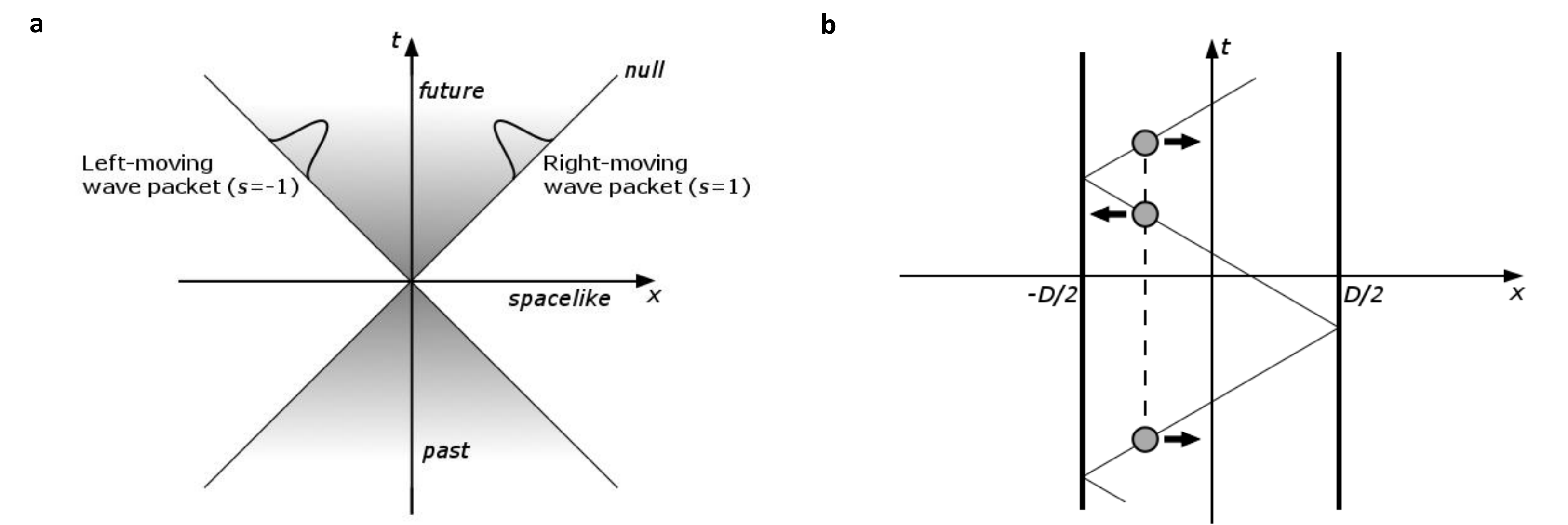}  
\caption{{\bf a.} Illustration of the two distinct types of dynamics of local wave packets of the quantised EM field. In the case of light propagation in one dimension, an initial wave packet of any shape can move in the direction of the positive and in the direction of the negative $x$ axis. When travelling at the speed of light it follows one of two null geodesics along the boundary of the light-cone. {\bf b.} Spacetime diagram depicting the dynamics of blips (represented by spheres) inside an optical cavity. Local excitations now pass through the same position many times.}
\label{fig:lightcone}
\end{figure*}

The purpose of this paper is two-fold. Firstly, we support the validity of a local field quantisation approach \cite{Jake} using ideas from special relativity. As we shall see below, it allows us to account for the two distinct types of dynamics of wave packets which travel along the $x$ axis: As illustrated in Fig.~\ref{fig:lightcone}(a), a wave packet of any shape can move either to the left or to the right. Secondly, we apply our approach to the modelling of the quantised EM field between two parallel mirrors and re-derive the Casimir effect for light propagation in one dimension. Instead of imposing the boundary condition of vanishing electric field amplitudes on the mirror surface by restricting the EM field inside the cavity to  standing waves with a discrete set of allowed frequencies, we realise boundary conditions in a dynamical fashion. As we shall see below, the sources of EM field amplitudes are local particles which are reflected whenever they come in contact with a mirror surface. As predicted by the mirror image method of classical electrodynamics, the resulting electric field amplitudes vanish always on the mirror surface. As we shall see below, this paper provides a more intuitive view on the Casimir effect, while also emphasising the importance of this effect for probing fundamental concepts in relativistic quantum field theories.  

\section{Results}

\subsection{A local relativistic free space theory}

Special relativity stipulates that the worldline of a photon moving in free space at the speed of light $c$ is a null-geodesic such that the spacetime interval ${\rm d} x^2 - c^{2} \, {\rm d} t^2  = ({\rm d} x + c \, {\rm d} t)({\rm d} x - c \, {\rm d} t) =0$. Hence the generators for the local excitations of the EM field must move along the same spacetime trajectory. As pointed out in Ref.~\cite{Jake}, this implies the following equation of motion,
\begin{eqnarray}  \label{eqn:EOM}
a_{s\lambda}(x,t) &=& a_{s\lambda}(x-s ct,0) \, , 
\end{eqnarray}
where the $a_{s \lambda} (x,t)$ is the annihilation operator for a basic energy quantum of light with spacetime coordinates $(x,t)$ and parameters $s$ and $\lambda$. From a quantum optics perspective, the $a_{s \lambda} (x,t)$ operators are local photon annihilation operators of the quantised EM field in the Heisenberg picture.

As illustrated in~Fig.~\ref{fig:lightcone}(a), independent of its initial shape, a wave packet propagating along the $x$ axis has two distinct orientations of its electric field vectors and two distinct directions of propagation: it has two polarisations and can move to the left or to the right. The resulting four-fold degeneracy is accounted for in the above notation by the parameters $s = \pm 1$ and $\lambda = {\sf H}, {\sf V}$. In contrast to quantum optics, relativistic quantum field theories already recognised the need to accommodate these independent degrees of freedom \cite{Dirac,Dirac2}. Hence it is not surprising that Eq.~(\ref{eqn:EOM}), when written as a first-order differential equation, 
\begin{eqnarray} \label{2}
\left( \frac{d}{dt}+s c\, \frac{d}{dx} \, \right) a_{s\lambda }(x,t) &=& 0, 
\end{eqnarray} 
has many similarities with the Dirac equation when simplified to the case of massless particles \cite{BB2}. In the above $s$ parametrises one of the two null world-lines and $x$ and $t$ can be any inertial space and time coordinates.  Notice also that this equation is valid in any reference frame since the speed of light is always the same.

The state vectors $|\psi(x,t) \rangle$ which span the total Hilbert space $\mathcal{H}$ of the quantised EM field in $1+1$ dimension are obtained by applying the creation operators $a^\dagger_{s \lambda}(x,t)$ repeatedly onto the vacuum state $|0 \rangle$ with $a_{s \lambda}(x,t) |0 \rangle = 0$. From Fig.~\ref{fig:lightcone}(a) we see that, in the absence of local interactions, spacetime-localised field excitations that have the same amplitude and belong to the same null-geodesic must be indistinguishable and must therefore have the same state vector. Moreover, states that describe excitations moving along different worldlines must be pairwise orthogonal. In the following, we ensure this by imposing
\begin{eqnarray}
\label{eqn:freespace_overlap}
\braket{1_{s\lambda}(x,t)|1_{s^\prime\lambda^\prime}(x^\prime,t)} 
&=& \langle 0| \big[ a_{s \lambda} (x,t) , a^\dagger_{s^{\prime }\lambda^{\prime }} (x^{\prime },t) \big] |0 \rangle ~~ \nonumber \\
&=& \delta_{ss^\prime} \, \delta_{\lambda\lambda^\prime} \, \delta(x-x^\prime) 
\end{eqnarray}
on the single-excitation states $|1_{s\lambda}(x,t) \rangle = a^\dagger_{s \lambda}(x,t) |0 \rangle$. The equivalence in the first line of the above equation shows that the $a_{s \lambda} (x,t)$ operators obey bosonic commutation relations, like the annihilation operators of the quantised EM field in momentum space \footnote{While the above overlap condition is intuitive, it is worth noting that it can also be derived using the Fourier transform while imposing the usual momentum-state commutator $[a_{s\lambda}(k,t),a^\dagger_{s^\prime\lambda^\prime}(k^\prime,t)]= \delta_{ss^\prime}\delta_{\lambda\lambda^\prime}\delta(k-k^\prime)$. By making use of the equation of motion in Eq.~(\ref{eqn:EOM}), we can modify the general commutator into an equal-time commutator before applying the Fourier transform.}. Due to the bosonic nature of the single excitations with states $|1_{s\lambda}(x,t) \rangle$, we refer to them in the following as {\em bosons localised in position} (blips) \cite{Jake}.

\begin{figure*}[t]
\centering
\includegraphics[width=\textwidth]{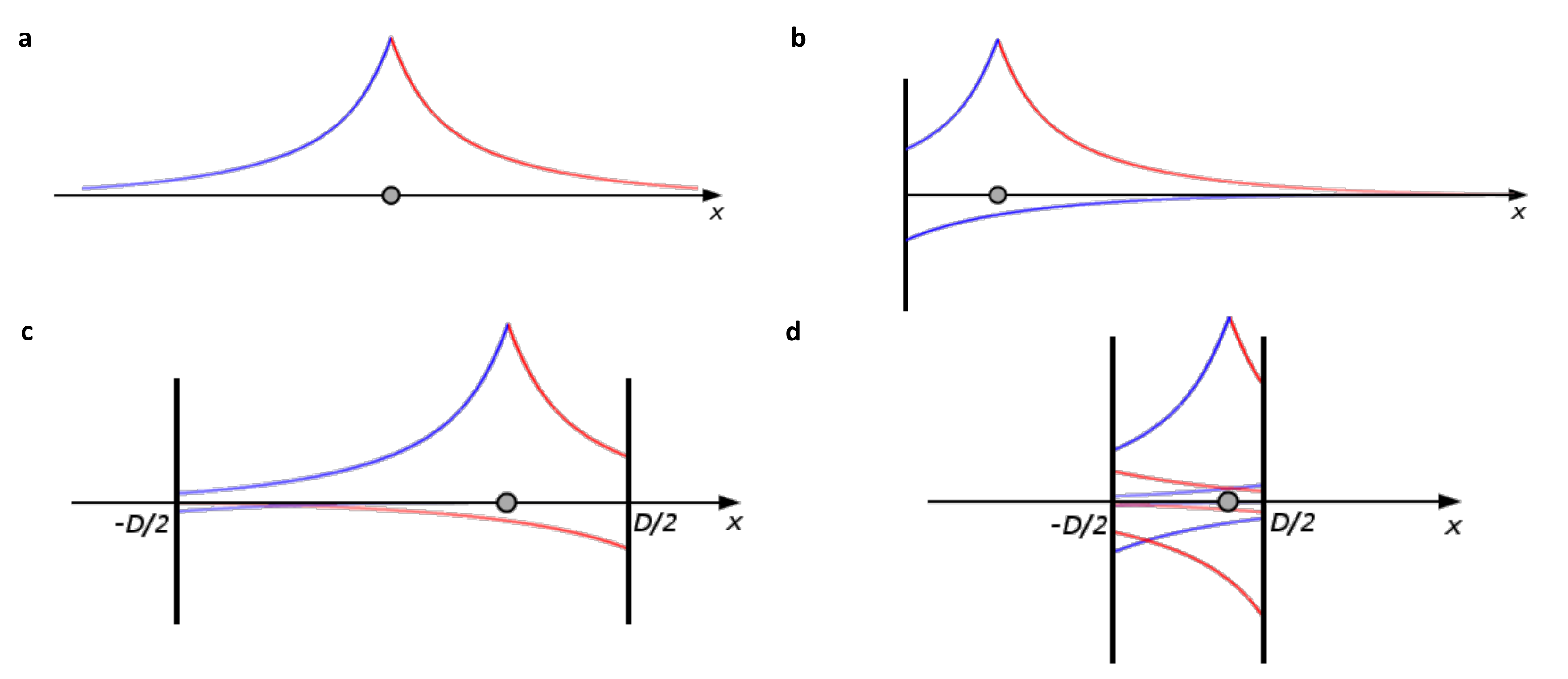}  
\caption{{\bf a.} Because of the regularisation operator ${\cal R}$ in Eq.~(\ref{EB}), local blip excitations contribute to local electric and magnetic field expectation values everywhere along the $x$ axis (cf.~Eq.~(\ref{EBlast})). {\bf b.} Since a blip on one side of a highly reflecting mirror cannot contribute to the field expectation value on the other side, its field contribution must be folded back on itself. This effect alters the electric and magnetic field observables in the presence of a mirror. {\bf c.} In the presence of two highly reflecting mirrors, blips outside the cavity cannot contribute to field expectation values on the inside. Moreover, the field contributions of blips on the inside need to be folded as in the case of one mirror. Now, however, the field contributions must be folded infinitely many times (cf.~Eq.~(\ref{EBcav2}) in Methods). {\bf d.} Comparing two cavities of different sizes, we see that the behaviour of the field contribution is now dependent on the cavity width. } \label{fig:1}
\end{figure*}

For many practical applications, like the modelling of the interaction of
the quantised EM field with local stationary objects, it is useful to write
Eq.~(\ref{eqn:EOM}) in the form of a Schr\"odinger equation,
\begin{eqnarray}  \label{eqn:SE}
\mathrm{i} \hbar \, |\dot \psi (x,t) \rangle &=& H_{\mathrm{dyn}}(t) \, |\psi (x,t) \rangle 
\end{eqnarray}
with $H_{\mathrm{dyn}}$ being the
relevant Hamiltonian to which interaction terms can be added.
A closer look at Eq.~(\ref{2}) which is a first-order linear differential equation containing a
time derivative shows that this is indeed possible. All $H_{\mathrm{dyn}}(t)$ needs to implement when used to generate the time evolution of an $a_{s \lambda} (x,t)$ operator is a spatial derivative. As recently shown in Ref.~\cite{Jake}, 
\begin{eqnarray}  \label{Hdyn222}
H_{\text{dyn}} (t) &=&  {1 \over 2\pi} \sum_{s, \lambda} \int_{-\infty}^{\infty}\text{d}x^{\prime
}\int_{-\infty}^{\infty}\text{d}x^{\prime \prime }\int_{-\infty}^\infty 
\mathrm{d} k \notag \,  \hbar c k \\
&&  \mathrm{e}^{\mathrm{i}s k (x^{\prime \prime
}-x^{\prime })} \, a^\dagger_{s\lambda}(x^{\prime \prime },t)
a_{s\lambda}(x^{\prime },t) 
\end{eqnarray}
has exactly this effect (cf.~Methods for more details). As the generator of the dynamics of photonic wave packets, the above Hamiltonian annihilates field excitations at positions $x^{\prime}$ and places them at positions $x^{\prime \prime }$ such
that excitations travel at the speed of light in their respective direction of
motion. It only assumes the form of a harmonic oscillator in the momentum-space
representation where it has positive and negative eigenvalues \cite{Jake}.  Moreover, one can show that the dynamical Hamiltonian is effectively time-independent and conserves energy.

From Maxwell's equations, we know that the electric and magnetic field expectation values $\langle \boldsymbol{E}(x,t) \rangle$ and $\langle \boldsymbol{B}(x,t) \rangle$ also propagate at the speed of light. Hence $\boldsymbol{E}(x,t) $ and $\boldsymbol{B}(x,t)$ must have the same spacetime dependence as $a_{s\lambda}(x,t)$ which suggests that 
\begin{eqnarray}  \label{EB}
\boldsymbol{E}(x,t) &=& \sum_{s= \pm 1} \mathcal{R} \left( a_{s\mathsf{H}}(x,t)\,%
\hat{\boldsymbol{y}} + a_{s\mathsf{V}}(x,t)\,\hat{\boldsymbol{z}} \right) + \mathrm{H.c.},  \notag
\\
\boldsymbol{B}(x,t) &=& \sum_{s= \pm 1} {\frac{s }{c}} \mathcal{R} \left( a_{s%
\mathsf{H}}(x,t)\,\hat{\boldsymbol{z}} - a_{s\mathsf{V}}(x,t)\,\hat{\boldsymbol{y}}
\right) + \mathrm{H.c.} ~~~~
\end{eqnarray}
Here $\hat{\boldsymbol{y}}$ and $\hat{\boldsymbol{z}}$ are unit vectors and $\mathcal{R}$ is a regularisation operator which does not depend on the spacetime coordinates $(x,t)$. As we shall see below, ${\cal R}$ imposes Lorentz covariance and determines the energy expectation value of single-blip excitations. Consistency with the classical expression for the energy in terms of $\boldsymbol{E}(x,t)$ and $\boldsymbol{B}(x,t)$ leads to (cf.~Methods)
\begin{equation}  \label{Heng}
H_{\text{energy}} =  \sum_{s, \lambda} \int_{-\infty}^{\infty}\text{d}x \, \varepsilon_0 A \left[ \,%
\mathcal{R}(a_{s\lambda}(x,t)) + \mathrm{H.c.} \, \right]^2 
\end{equation}
where $\varepsilon_0$ denotes the permittivity of free space and $A$ is the area which photons occupy in the $y$-$z$ plane when travelling along the $x$ axis. 

To determine ${\cal R}$, we take into account that spontaneous emission from an individual atom or ion results in the creation of exactly {\em one} photon. This assumption is in good agreement with quantum optics experiments which generate single energy quanta of light on demand \cite{Axelcav}. These behave as individual bosonic particles, acting as monochromatic waves with energies and frequencies determined by the atomic transition frequency $\omega_0$. As shown in Methods, $\mathcal{R}$ therefore adds a factor $N(k_0)$ to the momentum space ladder operators $a_{s \lambda}(k_0,t)$ and $a^\dagger_{s \lambda}(k_0,t)$ which relate to the blip annihilation and creation operators via complex Fourier transforms and establish Lorentz covariance of the electric and magnetic field operators. The above description therefore has many similarities with the standard description of the quantised EM field  in momentum space \cite{Sakurai,Loudon,Bennett}. However, the field operators in Eqs.~(\ref{EB}) and (\ref{Heng}) now act on a larger Hilbert space\textemdash positive and negative-frequency photons have been taken into account \cite{Jake}. They only transform into the standard expressions for $\boldsymbol{E}(x,t)$, $\boldsymbol{B}(x,t)$ and $H_{\text{energy}}$ when restricted to positive frequencies \cite{Jake}. In addition, we now characterise the local and the non-local excitations of the EM field not only by their positions $x$ and their wave numbers $k$ but also by their time of existence $t$. 

As shown in Methods, when returning from momentum to position space, we find that the electric and magnetic field observables now equal
\begin{eqnarray} \label{EBlast}
\boldsymbol{E}(x,t) &=& \sum_{s= \pm 1} \int^{\infty}_{-\infty} {\rm d}x' \left({\hbar c \over 16 \pi \varepsilon_0 A} \right)^{1/2} g(x,x') \nonumber \\
&& \left[ a_{s\mathsf{H}}(x',t)\,\hat{\boldsymbol{y}} + a_{s\mathsf{V}}(x',t)\,\hat{\boldsymbol{z}} \right] + \mathrm{H.c.} \, ,  \notag \\
\boldsymbol{B}(x,t) &=& \sum_{s= \pm 1}  \int^{\infty}_{-\infty} {\rm d}x' \, {s \over c} \left({\hbar c \over 16 \pi \varepsilon_0 A} \right)^{1/2} g(x,x') \nonumber \\ 
&& \left[ a_{s\mathsf{H}}(x',t)\,\hat{\boldsymbol{z}} - a_{s\mathsf{V}}(x',t)\,\hat{\boldsymbol{y}} \right] + \mathrm{H.c.} 
\end{eqnarray}
with the factor $g(x,x')$ given by \footnote{The last line in this equation holds for $x \neq x'$. For $x=x'$, the constant $g(x,x')$ diverges.}
\begin{eqnarray} \label{EBlast2}
g(x,x') &=& \int^{\infty}_{-\infty} {\rm d}k \left({2|k| \over \pi} \right)^{1/2} {\rm e}^{{\rm i} k(x-x')} \nonumber \\
&=& - |x-x'|^{-3/2} \, .  
\end{eqnarray}
Because of the presence of the superoperator ${\cal R}$ in Eq.~(\ref{EB}), local field expectation values contain contributions from blip excitations everywhere along the $x$ axis. Conversely, as illustrated in Fig.~\ref{fig:1}(a), local blip excitations contribute to electric and magnetic field expectation values everywhere along the $x$ axis. The commutator between $a_{s \lambda} (x',t)$ and $\boldsymbol{E}(x,t)$, for example, vanishes rapidly as the distance $|x-x'|$ increases, making this non-local effect small. However, it is not negligible and, as we shall see below, the non-locality of electric and magnetic field observables constitutes the origin of the Casimir effect. 

\subsection{Optical cavities and the Casimir effect}

When placed between the mirrors of an optical cavity, blips are continually reflected back and forth. As illustrated in Fig.~\ref{fig:lightcone}(b), they move on closed trajectories and travel through the same location $x$ many times.  Although the blips change direction when met with either of the mirrors, between the cavity walls they propagate freely.  Therefore, blips can be used to describe the EM field both in the absence \textit{and} in the presence of an optical cavity.  However, in order to capture their changed behaviour, we must replace the free space equation of motion in Eq.~(\ref{eqn:EOM}) by an alternative constraint. The dynamics of blips approaching the cavity walls can be described, for example, by a locally acting mirror Hamiltonian \cite{Jake}.  Another possibility to obtain blip operators which move along folded worldlines is to take inspiration from the mirror image method of classical electrodynamics \cite{Nick} and to map the dynamics of blips onto analogous free space scenarios.  

By adopting a local description, it is tempting to assume that the field expectation values of blips that are not in contact with the cavity do not depend on the presence or absence of highly reflecting mirrors at a spatially removed location. However the local electric and magnetic field observables ${\boldsymbol E}(x,t)$ and ${\boldsymbol B}(x,t)$ are {\em not} the same inside an optical cavity and in free space. As we have seen above (cf.~Fig.~\ref{fig:1}(a)), in free space, local blip excitations contribute to field expectation values everywhere along the $x$ axis. When constructing field observables in the presence of an optical cavity, we must take into account that its mirrors shield the inside of the cavity from light sources on the outside. We must therefore ensure that blips on the outside of the cavity do not contribute to electric and magnetic fields inside (Fig.~\ref{fig:1}(b)). Analogously, we must ensure that blips on the inside no longer contribute to fields on the outside. 

Here we are especially interested in highly reflecting mirrors with an amplitude reflection rate $r=-1$ with no light entering the cavity from the outside and no leakage of light out of the resonator.  We then hypothesise that the free space field amplitude contributions of local blips at positions $x$ with $x \in (-D/2,D/2)$ to local field observables beyond the cavity mirrors are reflected back where they contribute only to local field observables on the inside. More concretely, as illustrated in Figs.~\ref{fig:1}(b)-(d), when in contact with one of the mirror surfaces at positions $x= \pm D/2$, the field amplitudes of $a_{s \lambda}(x,t)$ blip excitations change their direction of propagation and start to contribute to 
the $(-s, \lambda)$ terms of the field observables. In addition, we need to take into account that electric field amplitudes accumulate a minus sign upon reflection. Hence the field observables $\boldsymbol{O}^{(\rm in)}_{s \lambda}(x,t)$ with $\boldsymbol{O} = \boldsymbol{E}, \boldsymbol{B}$ inside the resonator now equal
\begin{eqnarray}
	\label{field observables 2kkknewcav}
	\boldsymbol{O}^{(\rm in)}_{s \lambda}(x,t) &=& \sum_{n=-\infty}^\infty {\cal X} \left(\boldsymbol{O}^{(\rm free)}_{s \lambda}(x + 2nD, t) \right. \nonumber \\
	&& \left. - \boldsymbol{O}^{(\rm free)}_{-s \lambda}(-x + (2n-1)D, t) \right) 
\end{eqnarray}
where $\boldsymbol{O}^{(\rm free)}_{s \lambda}(x, t)$ denotes local free space observables and where the superoperator ${\cal X}$ restricts the Hilbert
space of the quantised EM field inside the cavity to local blip
excitations at positions $x \in (-D/2, D/2)$. The superoperator
does this by mapping each blip operator inside the cavity onto itself, and each
outside the cavity onto the zero operator. This thereby ensures that
$\boldsymbol{E}^{(\rm in)}(x,t) $ and $\boldsymbol{B}^{(\rm in)}(x,t) $ do not contain free space contributions which originate from blips on the outside (cf.~Eq.~(\ref{EBcav2}) in Methods).

As we have seen above, the reflections of the electric and magnetic field contributions of blips inside the cavity alter the electric and magnetic field observables in the presence of an optical cavity. As shown in Methods, the result is interference effects which reduce the zero point energy $H_{\rm ZPE}$ of the quantised EM field, thereby resulting in the Casimir attractive force 
\begin{equation} \label{final}
F_{\rm Casimir} = - {{\rm d} H_{\rm ZPE} \over {\rm d}D} = -{\pi \hbar c \over 6 D^2} 
\end{equation}
between the cavity mirrors, which is inversely proportional to the squared mirror distance $D^2$. As one would expect from comparing Figs.~\ref{fig:1}(c) and (d), the smaller the amount of interference within the cavity, the smaller the resulting Casimir force. Our approach also demonstrates that the change of the local field observables on the outside of the cavity, which we illustrate in Fig.~\ref{fig:1}(b), does not contribute to the Casimir force (cf.~Methods for more details).

In contrast to previous derivations of Casimir forces \cite{Casimir,interest}, our analysis singles out the finite $D$-dependent term from the $D$-independent divergent contributions to the zero point energy $H_{\rm ZPE}$ without the need for explicit regularisation prescriptions. Here we attribute the above Casimir force to a change of the topology of the electric and magnetic field observables associated with blip excitations {\em inside} the cavity (with no contributions from {\em external} blips) which ensures zero electric field boundary conditions on the mirror surface. Notice also that our result differs by a factor of four from the results of previous authors \cite{Bordag} since we consider two polarisation degrees of freedom (not only one) and allow for positive and negative frequency photons.  Similarly, the zero point energy in free space which we derive in Methods contains a factor of two (cf.~Eq.~(\ref{ZPE})). 

\section{Discussion}

Compared to its standard description, we parametrise the Hilbert space of the quantised EM field by treating space {\em and} time equivalently, and model the dynamics of its states in terms of a Hamiltonian constraint. This constraint ensures that local field excitations belonging to the same worldline have the same bosonic field annihilation operators $a_{s \lambda}(x,t)$. In free space, this approach is shown to be analogous to describing the dynamics of quantum states of light by a Schr\"odinger equation, but requires a dynamical Hamiltonian with positive and negative eigenvalues which no longer coincides with the energy observable. Moreover, we find it useful to distinguish between the local building blocks of light, blips, and the field that they create. As illustrated in Fig.~\ref{fig:1}(a), in free space, local blip excitations contribute to electric and magnetic field expectation values everywhere along the $x$ axis. In this sense, blips are localised particles which do not carry mass nor charge but constitute the sources of non-local electric magnetic fields. They are thus similar to local charged particles which create nonlocal electric field amplitudes and to massive particles which create nonlocal gravitational fields.

When extending our approach to the modelling of light scattering in optical cavities, one must therefore ensure that blips outside the cavity do not contribute to local energy densities inside and vice versa. This has implications for the form of the electric and magnetic field observables inside an optical cavity (cf.~Eq.~(\ref{EBcav2})) but these can now be written in terms of the free space field annihilation and creation operators. When applied to the Casimir effect, a local relativistic description provides additional insight by attributing its force to the change of the topology of the quantised EM field inside the cavity. The methodology which we introduced in this paper, once extended to light propagation in three dimensions, can be adjusted to study Casimir forces in a more straightforward way in a wide range of situations involving different geometries, moving mirrors and so on, while also taking actual material constants, like mirror reflection rates  and finite temperatures, into account. In addition, our approach emphasises that optical cavities support a continuum of photon frequencies which is important for the modelling of Fabry-Perot cavities \cite{Barlow} and in good agreement with recent experiments with nanocavities that confine light to tight spaces which are much smaller than optical wavelengths \cite{Baumberg}. 

\section{Methods}

\subsection{Consistency of Eqs.~(\protect\ref{2}) and (\ref{Hdyn222})} \label{app_consistency}

Any operator $O(x,t)$ in the Heisenberg picture satisfies the Heisenberg equation $\dot O(x,t) = -{\rm i}/\hbar \, [O(x,t), H_{\rm dyn}]$.
Suppose $O(x,t) = a_{s\lambda}(x,t)$ and $H_\text{dyn}$ is the dynamical Hamiltonian specified in Eq.~(\ref{Hdyn222}). Using the commutation relation in Eq.~(\ref{eqn:freespace_overlap}), we therefore find that  
\begin{eqnarray}
\dot a_{s\lambda}(x,t) 
&=& -\frac{{\rm i}c}{2\pi} \int_{-\infty}^{\infty}\text{d}x' \int_{-\infty}^{\infty}\text{d}k \, k \, {\rm e}^{{\rm i}s k(x-x')} \, a_{s\lambda} (x',t)\nonumber\\
&=& -s c \, \frac{\text{d}}{\text{d}x} \, \int_{-\infty}^{\infty}\text{d}x' \, \delta(x-x') \, a_{s\lambda}(x',t) \nonumber\\
&=& -s c \, \frac{\text{d}}{\text{d}x} \, a_{s\lambda}(x,t)
\end{eqnarray}
which shows the consistency of Eqs.~(\ref{2}) and (\ref{Hdyn222}).

\subsection{The superoperator ${\cal R}$ and the zero-point energy $H_{\rm ZPE}$ in free space} \label{newappB}

The energy observable of the quantised EM field in free space can be derived from its classical expression in terms of electric and magnetic fields,
\begin{eqnarray} \label{Hengclass}
H_{\text{energy}} &=& {A \over 2} \int_{-\infty}^{\infty}\text{d}x \, \Big[ \varepsilon_0 \boldsymbol{E}(x,t)^2 + \frac{1}{\mu} \boldsymbol{B}(x,t)^2 \Big] . ~~~
\end{eqnarray}
Substituting the field observables in Eq.~(\ref{EB}) in terms of ladder operators into this equation leads to Eq.~(\ref{Heng}), which contains the superoperator ${\cal R}$. To evaluate this expression, we replace the position-space operators by their Fourier transforms \cite{Jake},
\begin{eqnarray} \label{field observables 2kkknew}
a_{s \lambda}(x,t) &=& {1 \over \sqrt{2 \pi}} \int^{\infty}_{-\infty} {\rm d}k \, {\rm e}^{{\rm i}s kx} \, a_{s \lambda}(k,t) \, ,
\end{eqnarray}
where the $a_{s \lambda}(k,t)$ are bosonic momentum space annihilation operators with $[ a_{s\lambda}(k,t), a^\dagger_{s'\lambda'}(k',t) \big] = \delta_{\lambda\lambda'}\delta_{ss'}\delta(k-k')$. Assuming that the superoperator ${\cal R}$ multiplies $a_{s \lambda}(k,t)$ with a (real) factor $N(k)$ which is independent of $s$, $\lambda$ and $t$ for symmetry reasons, and taking into account that symmetry implies $N(k) = N(-k)$, we then find that  
\begin{eqnarray}  \label{Heng2}
	H_{\text{energy}} &=& \varepsilon_0 A \sum_{s, \lambda} \int_{-\infty}^{\infty}\text{d}k \, N(k)^2 \, \big[ a^\dagger_{s\lambda}(k,t)a_{s\lambda}(k,t) \nonumber \\
		&& + a_{s\lambda}(k,t)a_{s\lambda}(-k,t) + {\rm H.c.} \big]  ~~~
\end{eqnarray}
which is always positive. Suppose a single two-level atom with transition frequency $\omega_0 = c |k_0|$ which is initially in its excited state creates exactly one photon after interacting for some time with the free radiation field. Due to the resonance of the atom-field interaction, the frequency of this photon must be the same as that of the atom. Moreover, due to energy conservation, its energy must be the same as the initial energy of the atom. Hence a photon with wave number $k_0$ must have the energy $\hbar c |k_0|$ which leads us to $N(k_0) = ( \hbar c  |k_0| / 2 \varepsilon_0 A)^{1/2}$. Hence, the regularisation operator ${\cal R}$ simply multiplies the momentum-space operators $a_{s \lambda}(k)$ and $a^\dagger_{s \lambda}(k)$ with a factor proportional to $|k|^{1/2}$. Next let us have a closer look at the implications of the above calculations for the electric and magnetic field observables $\boldsymbol{E}(x,t)$ and $\boldsymbol{B}(x,t)$. Substituting Eq.~(\ref{field observables 2kkknew}) into Eq.~(\ref{EB}), applying the regularisation operator ${\cal R}$ and employing the inverse Fourier transform
\begin{eqnarray} \label{field observables 2kkk}
a_{s \lambda}(k,t) &=& {1 \over \sqrt{2 \pi}} \int^{\infty}_{-\infty} {\rm d}x' \, {\rm e}^{-{\rm i}s kx'} \, a_{s \lambda}(x',t) \, ,
\end{eqnarray}
we obtain Eq.~(\ref{EBlast}) in the main text. Finally, we calculate the zero-point energy  $H_{\rm ZPE} =\braket{0|H_\text{energy}|0} $ of the quantised EM field in free space. From Eq.~(\ref{Heng}) we see that this vacuum expectation value equals 
\begin{eqnarray} \label{ZPE}
H_{\rm ZPE} &=& {\hbar c \over 2\pi} \int_{-\infty}^{\infty}\text{d}x \int_{-\infty}^{\infty}\text{d}k \, |k| 
\end{eqnarray}
which is infinitely large. The reason for this is that $H_{\rm ZPE}$ is not only a vacuum expectation value,
it is also a measure of the size of the single-excitation Hilbert space of the quantised EM field in free space. Our result therefore differs from the standard result by a factor two, reflecting that the Hilbert space of the quantised EM field in free space has been doubled in this paper.

\subsection{The zero-point energy $H_{\rm ZPE}$ in the presence of an optical cavity} \label{app_cavity}

When constructing the electric and the magnetic field observables $\boldsymbol{E}^{(\rm in)}(x,t)$ and $\boldsymbol{B}^{(\rm in)}(x,t)$ of the quantised EM field inside an optical cavity, we need to reflect the free space contributions of local blips on the outside of the cavity back inside. In addition, we need to remove all contributions from blips located on the outside of the cavity to the quantised EM field on the inside as suggested in Eq.~(\ref{field observables 2kkknewcav}). By substituting Eq.~(\ref{EBlast}) into this equation, we find that the electric and magnetic field observables inside the cavity are given by
\begin{eqnarray} \label{EBcav2}
\boldsymbol{E}^{(\rm in)}(x,t) &=& -\sum_{n=-\infty}^\infty \sum_{s= \pm 1} \int^{D/2}_{-D/2} {\rm d}x'  \left({\hbar c \over 16\pi\varepsilon_0 A} \right)^{1/2} \notag \\
&& \hspace*{-1.3cm} \left[ \left|x-x'+2nD \right|^{-3/2} - \left| x+x' +(2n-1)D \right|^{-3/2} \right] \nonumber \\
&& \hspace*{-1.3cm} \times \left[ a_{s\mathsf{H}}(x',t)\,\hat{\boldsymbol{y}} + a_{s\mathsf{V}}(x',t)\,\hat{\boldsymbol{z}} \right] + \mathrm{H.c.} \, , \notag \\
\boldsymbol{B}^{(\rm in)}(x,t) &=&- \sum_{n=-\infty}^\infty \sum_{s= \pm 1} \int^{D/2}_{-D/2} {\rm d}x'  \, {s \over c} \left({\hbar c \over 16\pi\varepsilon_0 A} \right)^{1/2} \notag \\ 
&& \hspace*{-1.3cm} \left[ \left| x-x'+2nD \right|^{-3/2} + \left| x + x' +(2n-1)D \right|^{-3/2} \right] \notag \\
&& \hspace*{-1.3cm} \times \left[ a_{s\mathsf{H}}(x',t)\,\hat{\boldsymbol{z}} - a_{s\mathsf{V}}(x',t)\,\hat{\boldsymbol{y}} \right] + \mathrm{H.c.} 
\end{eqnarray}
These equations express the field observables inside the cavity as position-dependent superpositions of the bosonic blip operators inside the cavity. 

Next we obtain an expression for the observable of the energy within the cavity by substituting the above field observables into Eq.~(\ref{Hengclass}) but with the $x$ integration being carried out over the width of the cavity only. Using the bosonic commutation relations in Eq.~(\ref{eqn:freespace_overlap}) and performing one position integral, one can then show that the zero point energy of the quantised EM field inside the cavity equals   
\begin{eqnarray}\label{ZPE1}
H_{\text{ZPE}}^{\text{(in)}} &=& \frac{\hbar c}{4 \pi} \sum_{n,m = -\infty}^{\infty} \int_{-D/2}^{D/2}\text{d}x \int_{-D/2}^{D/2}\text{d}x' \notag \\
&& \hspace*{-0.8cm} \left[  \left|(x+ x' + (2n-1)D)(x+ x' + (2m-1)D) \right|^{-3/2} \right. \nonumber \\
&& \hspace*{-0.8cm} \left. + \left|(x- x' +2nD )(x- x' +2mD) \right|^{-3/2} \right] \, .
\end{eqnarray}
The first term in this expression can be made to look like the second term, apart from different integral limits, by substituting $\tilde x' = - x'$; $\tilde x = x + D$, $\tilde n = n -1$ and $\tilde m = m - 1$ when $x<0$; and $\tilde x = x - D$ when $x>0$. Doing so, we find that the total zero-point energy of the quantised EM field inside the cavity equals
\begin{widetext}
\begin{eqnarray}\label{ZPE2}
H_{\text{ZPE}}^{\text{(in)}} &=& \frac{\hbar c}{4 \pi} \sum_{n,m = -\infty}^{\infty} \int_{-D}^{D}\text{d}x \int_{-D/2}^{D/2}\text{d}x' \, \left|(x- x' + 2nD)(x- x' + 2mD) \right|^{-3/2} \nonumber \\
&=& \frac{\hbar c}{4 \pi} \sum_{n,m = -\infty}^{\infty} \int_{-D+2nD}^{D+2nD}\text{d}x \int_{-D/2}^{D/2}\text{d}x' \, \left| ( x- x' ) (x- x' + 2(m-n)D) \right|^{-3/2} \nonumber \\
&=& \frac{\hbar c}{4 \pi} \sum_{m = -\infty}^{\infty} \int_{-\infty}^\infty \text{d}x \int_{-D/2}^{D/2}\text{d}x' \, \left|(x- x')(x- x' + 2mD) \right|^{-3/2} \, .
\end{eqnarray}
\end{widetext}
The latter applies since the term $m-n$ takes all values between $-\infty$ and $\infty$ as we sum over $m$ irrespective of $n$ and since the sum over $n$ in the first line of this equation has the effect of extending the $x$ integral from the range $(-D,D)$ to $(-\infty, \infty)$. Since it can be shown that
\begin{eqnarray}
	\label{ZPE4}
	\int_{-\infty}^{\infty}\text{d}x \, |(x_1-x)(x-x_2)|^{-3/2} &=& - {8 \over |x_1-x_2|^2} \, , ~~
\end{eqnarray}
the vacuum energy inside the resonator equals
\begin{eqnarray}
	\label{ZPE5}
	H_{\text{ZPE}}^{\text{(in)}} &=& -\frac{\hbar c}{2\pi D}\sum_{m=-\infty}^{\infty}\frac{1}{m^2}\, .
\end{eqnarray}
The $m=0$ contribution in this equation is divergent, but it can be calculated by returning to the first line in Eq.~(\ref{EBlast2}) and Eq.~(\ref{ZPE2}), which show that the $D$-dependence of this term is linear. That is, the energy density due to this term is constant. Furthermore, its value is identical to the contribution to the zero point energy of an equal-sized region of free space. We also need to consider the zero point energy $H_{\text{ZPE}}^{\text{(out)}}$ of the EM field outside the cavity mirrors. This contribution to the total zero point energy $H_{\text{ZPE}}$ can be calculated analogously by taking into account the reflection of field contribution on the outside of the cavity mirrors, as illustrated in Fig.~\ref{fig:1}(b). Again, the resulting external contribution is identical to that of an equally sized region of free space. As such, the contributions of both external regions, together with the $m=0$ term reproduce the full free space zero point energy. Consequently, the $m=0$ term does not contribute to the Casimir force which we present in Eq.~(\ref{final}). To arrive at this force, we require the Basel sum which states that $\sum_{m = 1}^{\infty} m^{-2} = \pi^2/6$.

\section*{\small{Data Availability}}

\noindent Statement of compliance with EPSRC policy framework on research data: This publication is
theoretical work that does not require supporting research data.

\section*{\small{Acknowledgements}}

\noindent A.B., D.H. and R.P. would like to thank Nicholas Furtak-Wells, Jiannis Pachos and Jake Southall for stimulating discussions. D. H. acknowledges financial support from the UK Engineering and Physical Sciences Research Council EPSRC.

\section*{\small{Author contributions}} 

\noindent All authors contributed to the theoretical modelling, the understanding of the results and the writing of the manuscript.  

\section*{\small{Competing interests}}

The authors declare no competing interests.

\end{document}